\documentclass[final]{article}
\usepackage{physics}
\usepackage[final]{graphicx}
\usepackage{graphicx}
\usepackage{amssymb}
\usepackage{rotating}
\usepackage{pdflscape}
\usepackage[english]{babel}
\usepackage{color}
\usepackage{bigintcalc}
\def\beq{\begin{eqnarray}}
\def\eeq{\end{eqnarray}}
\usepackage{amsfonts}

\usepackage{seqsplit}
\newtheorem{rule-of-thumb}[theorem]{Definition} % Definite {rule-of-thumb}

\begin{document}

\title{Comment to: "The quantum square well with moving boundaries: a numerical analysis"}

\author{Miguel Ahumada-Centeno \\
Facultad de Ciencias, CUICBAS, Universidad de Colima,\\
Bernal D\'{i}az del Castillo 340, Colima, Colima, Mexico  \\
mahumada0@ucol.mx \\
\and
Paolo Amore \\
Facultad de Ciencias, CUICBAS, Universidad de Colima,\\
Bernal D\'{i}az del Castillo 340, Colima, Colima, Mexico  \\
paolo.amore@gmail.com 
\and
Francisco M. Fern\'andez\\
INIFTA,  Division Quimica Teorica,  \\
Blvd. 113 S/N, Sucursal 4, Casilla de Correo 16, 1900\\
 La Plata, Argentina \\
 framfer@gmail.com}

\maketitle

\begin{abstract}
In this comment we show that the approach presented by Foj\'on et al~\cite{Fojon10} is not as accurate as they claim.
A straightforward calculation using the models considered buy those authors clearly shows that the spectral method, which the authors criticize,
proves to be considerably better.
\end{abstract}

%%%%%%%%%%%%%%%%%%%%%%%%%%%%%%%%%%%%%%%%%%%%%%%%%%%%%%%%%%%%%%%%%%%%%%%%%%%%%%%%%%%%%%%%%%%%%%%%%%%%%%%%%%%%%%%%%%%%%%%
\section{Introduction}
\label{sec:intro}
%%%%%%%%%%%%%%%%%%%%%%%%%%%%%%%%%%%%%%%%%%%%%%%%%%%%%%%%%%%%%%%%%%%%%%%%%%%%%%%%%%%%%%%%%%%%%%%%%%%%%%%%%%%%%%%%%%%%%%%

Ref.~\cite{Fojon10} discusses the numerical solution of the time-dependent Schr\"odinger equation (TDSE) in a one dimensional 
square well with moving boundaries. This problem is exactly solvable only in the special cases, such as for a wall moving 
at constant velocity, as first discussed by Doescher and Rice long time ago~\cite{Doescher69}.  Makowski and 
Dembinski~\cite{Makowski91} have proved that if $\ell''(t) \ell^3(t) = {\rm constant}$, where $\ell(t)$ describes the position 
of the right wall (the left wall being  fixed at the origin), then the problem is solvable in terms of suitable transformations of the hamiltonian.

In the majority of the cases, since no exact solution is available, one has to resort either to approximate techniques (such as
perturbation theory) or to numerical methods. Doescher and Rice~\cite{Doescher69}, for example, have used first order perturbation theory,  
to obtain the probability that a particle initially in the ground state may be found in the first excited state at later times,
when the right wall is moving. Unfortunately this approximation works well only when the wall moves slowly, but it fails at larger
velocities. Extending the perturbation calculation to higher orders, on the other hand, is not desirable for this problem, since it may involve
in any case numerical calculations and the complexity of the task will depend on the maximal order in perturbation theory that one
needs to take into account (leaving aside the issue of convergence of the perturbation expansion).

The numerical approach is therefore the optimal choice for the problem at hand, although different strategies may be adopted in
its implementation.  The approach adopted in Ref.~\cite{Fojon10} is based on a suitable scaling of the space variable, followed
by discretization of the space, which converts the original TDSE into a system of $n-1$ differential equations in the time variable
(eqs.(19) of Ref.~\cite{Fojon10}). 
A different approach, that we will follow in our analysis, consists of numerically solving the system of coupled first order 
differential equations for the coefficients of the decomposition of the exact solution in the basis of the instantaneous eigenfunctions 
of the well, as discussed in Ref.~\cite{Doescher69}. The approach followed  in Ref.~\cite{Fojon10} consists of a spatial discretization,
leading to a system of first order differential equations in the time variable, that are then solved by numerical integration; in the spectral approach,
on the other hand, the only approximation made is the truncation over the number of modes, provided that the relevant integrals 
can be performed exactly.
On these grounds, one should expect that the spectral method would be superior to the method of Ref.~\cite{Fojon10}.

Despite this observation, the authors of Ref.~\cite{Fojon10} claim a superiority of their approach on the spectral approach, concluding that 
"the (spectral) method is not only complicated, but, in addition, all these manipulations are often sources of errors, and therefore, the
final result is not accurate".

The purpose of this paper is to compare the two methods and provide the evidence that the method of Ref.~\cite{Fojon10} is in general 
much less accurate than the spectral method; additionally, we show that the results discussed in Ref.~\cite{Fojon10} for the cases of 
fast movements are incorrect and plagued by much larger errors than claimed by the authors.

The paper is organized as follows: in section \ref{sec:model} we briefly discuss the problem and obtain the relevant equations
to be solved; in section \ref{sec:num} we present our numerical results and compare them with the analogous results of Ref.~\cite{Fojon10}.
Finally, in section \ref{sec:concl} we state our conclusions.

\section{The model}
\label{sec:model}

Consider the time-dependent Schr\"odinger equation
\begin{eqnarray}
i \hbar \frac{\partial \psi}{\partial t} = \hat{H} \psi(x,t)
\label{eq_Sch}
\end{eqnarray}
where $\hat{H}$ is the Hamiltonian  for a particle in one dimension
confined in an infinite square well with a moving wall
\begin{eqnarray}
\hat{H} = - \frac{\hbar^2}{2 m} \frac{d^2}{dx^2} +  V(x,t)
\end{eqnarray}
and
\begin{eqnarray}
V(x,t) = \left\{
\begin{array}{ccc}
\infty & , & x\leq 0  \ \ , \ \ x \geq \ell(t) \\
0 & , & 0<x < \ell(t)
\end{array}
\right.
\end{eqnarray}

The wave function obeys the time-dependent Dirichlet boundary conditions
\begin{eqnarray}
\psi(0, t) = \psi(\ell(t),t) = 0
\end{eqnarray}

Let $u_n(x,t)$ be the instantaneous energy eigenfunctions
\begin{eqnarray}
u_n(x,t) = \sqrt{\frac{2}{\ell(t)}} \ \sin \frac{n\pi x}{\ell(t)}
\label{eq_iee}
\end{eqnarray}
and $E_n(t)$ be the instantaneous energy eigenvalues
\begin{eqnarray}
E_n(t) = \frac{\hbar^2\pi^2 n^2}{2m\ell^2(t)} \ .
\end{eqnarray}

These eigenfunctions form a basis and therefore one can expand a solution 
to the time-dependent Schr\"odinger equation in this basis as~\cite{Doescher69}
\begin{eqnarray}
\psi(x,t) =  \sum_n b_n(t) u_n(x,t) %\mathcal{F}_n(t)
e^{-\frac{i}{\hbar} \int_0^t E_n(\tau) d\tau} \ .
\end{eqnarray}

Using this expansion inside eq.~(\ref{eq_Sch}), one can use the orthonormality of $\left\{ u \right\}$
to obtain an infinite set of coupled first order differential equations for the expansion coefficients $b_k$ as
\begin{eqnarray}
\dot{b}_k(t) &=& - \sum_n b_n(t) \Delta_{kn}(t) e^{\frac{i}{\hbar} \int_0^t (E_k(\tau)-E_n(\tau)) d\tau} 
\label{eq_bs}
\end{eqnarray}
where we have introduced the definition
\begin{eqnarray}
\Delta_{kn}(t) &\equiv& \int_0^{\ell(t)} u_k(x,t) \frac{d}{dt} u_n(x,t) dx \nonumber \\
&=& \left\{
\begin{array}{ccc}
0 & , & k=n \\
(-1)^{k+n} \frac{2 kn}{k^2-n^2} \frac{\ell'(t)}{\ell(t)} &  , & |k-n|>0 \\
\end{array}
\right.
\end{eqnarray}

It is straightforward to see that eqs.(\ref{eq_bs}) preserve  the total probability $P = \sum_k |b_k|^2$.

Let us define
\begin{eqnarray}
M_{kn}(t) \equiv \Delta_{kn} e^{i \eta_{kn}(t)}
\end{eqnarray}
where $\eta_{kn}(t) \equiv {\frac{1}{\hbar} \int_0^t (E_k(\tau)-E_n(\tau)) d\tau}$.

It is easy to prove that ${\bf M}$ is antihermitian~\footnote{The conservation of total probability follows directly 
from this property.}
\begin{eqnarray}
{\bf M}^\dagger(t) = - {\bf M}(t)
\end{eqnarray}

The straightforward approach to the solution of eqs.~(\ref{eq_bs}) is to use the standard techniques for solving
systems of ODEs, such as the Runge-Kutta method. In this case it is preferable to work with real quantities and
we introduce the definitions
\begin{eqnarray}
c_k \equiv \Re(b_k) \hspace{0.5cm} , \hspace{0.5cm} d_k \equiv \Im(b_k) 
\end{eqnarray}
and express eqs.~(\ref{eq_bs}) into an equivalent set of coupled differential equations as
\begin{eqnarray}
\dot{c}_k(t) &=& - \sum_{n \neq k} \Delta_{kn}(t) \left[ c_n(t)  \cos\left( \eta_{kn}(t) \right)
- d_n(t)  \sin\left( \eta_{kn}(t) \right) \right] \label{eq_c} \\
\dot{d}_k(t) &=& - \sum_{n \neq k} \Delta_{kn}(t) \left[ c_n(t)  \sin\left( \eta_{kn}(t)\right)
+ d_n(t)  \cos\left( \eta_{kn}(t) \right) \right] \label{eq_d} \ .
\end{eqnarray}

To numerically solve these equations, one needs to restrict the calculation to a finite number of modes
by imposing the condition $k \leq k_{\rm MAX}$. The value of $k \leq k_{\rm MAX}$ should be chosen 
in such a way that the coefficients $c_k$ and $d_k$ are negligible for $k > k_{\rm MAX}$.

The initial conditions ($c_k(t_0) = c_k^{(0)}$ and $d_k(t_0) = d_k^{(0)}$) and the law specifying the 
movement of the right wall ($\ell(t)$) will also play a role in determining a suitable value of the
cutoff $k_{\rm MAX}$.

For the special case of a box with a uniformly moving wall, Doescher and Rice ~\cite{Doescher69}
have obtained the solution to eq.~(\ref{eq_Sch}) exactly in the form
\begin{eqnarray}
\Psi_n(x,t) =  \sqrt{\frac{2}{\ell(t)}} e^{i \alpha \xi (\frac{x}{\ell(t)})^2 - i \frac{n^2 \pi^2}{4\alpha} (1-1/\xi)} \ \sin (\frac{n\pi x}{\ell(t)}) 
\hspace{1cm} n=1,2,\dots
\label{eq_exact}
\end{eqnarray}
where $\xi(t) \equiv \ell(t)/\ell(0)$ and $\alpha \equiv \frac{m}{2\hbar} \ell(0) \frac{d\ell}{dt}$.

\section{Numerical solutions}
\label{sec:num}

We discuss the numerical solution of the eqs.~(\ref{eq_c}) and (\ref{eq_d}) for different laws of
motion of the right wall. The numerical results will be contrasted with the results of 
refs.~\cite{Doescher69,Fojon10}. 

Doescher and Rice~\cite{Doescher69} consider the time evolution of a particle that at the initial time is in
the ground state of the wall:
\begin{eqnarray}
\psi(x,0) = u_1(x,0) %\hspace{3cm} (Doescher \  and  \ Rice)
\label{eq_in_Doescher}
\end{eqnarray}
whereas Foj\'on et al.~\cite{Fojon10} adopt a less conventional choice
\begin{eqnarray}
\psi_j(x,0) = u_j(x,0) e^{i x^2 \frac{l'(0)}{4 \ell(0)}} %\hspace{3cm} (Foj\acute{o}n \ et \ al.)
\label{eq_in_Fojon}
\end{eqnarray}
corresponding to the $j$ state of the box with a uniformly moving wall at the initial time 
(notice that in the calculation of \cite{Fojon10}  $\hbar=1$ and $m=1/2$).

The initial wavefunction (\ref{eq_in_Fojon})  can be decomposed in the basis of the instantaneous eigenfunction
\begin{eqnarray}
\psi_j(x,0) = \sum_{k} q_{jk} u_k(x,0) 
\end{eqnarray}
where
\begin{eqnarray}
q_{jk} =  \int_0^{\ell(0)} \psi_j(x,0) u_k(x,0) dx 
\end{eqnarray}

The expansion coefficients $q_{jk}$ can be calculated exactly in terms of Fresnel functions
\begin{eqnarray}
\Re \left[q_{jk} \right] &=& \frac{\sqrt{\frac{\pi }{2}}}{2 \sqrt{\alpha }}
   \left[\left(C\left(\frac{\pi  j+2 \alpha -k \pi}{\sqrt{2 \pi } \sqrt{\alpha}}\right)
   -C\left(\frac{\pi  j-2 \alpha -k \pi}{\sqrt{2 \pi } \sqrt{\alpha }}\right)\right)
   \cos \left(\frac{\pi ^2 (j-k)^2}{4 \alpha}\right) \nonumber \right. \\
   &+& \left. \left(C\left(\frac{(j+k) \pi -2 \alpha}{\sqrt{2 \pi } \sqrt{\alpha}}\right)
   -C\left(\frac{\pi  (j+k)+2 \alpha}{\sqrt{2 \pi } \sqrt{\alpha }}\right)\right) 
   \cos \left(\frac{\pi ^2 (j+k)^2}{4 \alpha}\right) \nonumber \right. \\
   &+& \left. \left(S\left(\frac{\pi  j+2 \alpha -k\pi }{\sqrt{2 \pi } \sqrt{\alpha}}\right)
   -S\left(\frac{\pi  j-2 \alpha -k \pi}{\sqrt{2 \pi } \sqrt{\alpha }}\right)\right)
   \sin \left(\frac{\pi ^2 (j-k)^2}{4 \alpha}\right) \nonumber \right. \\
   &+& \left. \left(S\left(\frac{(j+k) \pi -2 \alpha
   }{\sqrt{2 \pi } \sqrt{\alpha
   }}\right)-S\left(\frac{\pi  (j+k)+2 \alpha
   }{\sqrt{2 \pi } \sqrt{\alpha }}\right)\right)
   \sin \left(\frac{\pi ^2 (j+k)^2}{4 \alpha
   }\right)\right] \nonumber \\
\Im \left[q_{jk} \right] &=& \frac{\sqrt{\frac{\pi }{2}}}{2 \sqrt{\alpha }}
   \left[\left(C\left(\frac{\pi  j-2 \alpha -k \pi}{\sqrt{2 \pi } \sqrt{\alpha}}\right)
   -C\left(\frac{\pi  j+2 \alpha -k \pi}{\sqrt{2 \pi } \sqrt{\alpha }}\right)\right)
   \sin \left(\frac{\pi ^2 (j-k)^2}{4 \alpha}\right) \right. \nonumber \\
   &+& \left. \left(C\left(\frac{\pi  (j+k)+2 \alpha}{\sqrt{2 \pi } \sqrt{\alpha}}\right)
   -C\left(\frac{(j+k) \pi -2 \alpha}{\sqrt{2 \pi } \sqrt{\alpha }}\right)\right) 
   \sin \left(\frac{\pi ^2 (j+k)^2}{4 \alpha}\right) \right. \nonumber \\
   &+& \left. \left(S\left(\frac{\pi  j+2 \alpha -k\pi }{\sqrt{2 \pi } \sqrt{\alpha}}\right)
   -S\left(\frac{\pi  j-2 \alpha -k \pi}{\sqrt{2 \pi } \sqrt{\alpha }}\right)\right)
   \cos \left(\frac{\pi ^2 (j-k)^2}{4 \alpha}\right) \right. \nonumber \\
   &+& \left. \left(S\left(\frac{(j+k) \pi -2 \alpha}{\sqrt{2 \pi } \sqrt{\alpha}}\right)
   -S\left(\frac{\pi  (j+k)+2 \alpha}{\sqrt{2 \pi } \sqrt{\alpha }}\right)\right)
   \cos \left(\frac{\pi ^2 (j+k)^2}{4 \alpha}\right)\right] \nonumber
\end{eqnarray}
where $C(x) \equiv  \int_0^x \cos \frac{\pi t^2}{2} dt$ and $S(x) \equiv  \int_0^x \sin \frac{\pi t^2}{2} dt$
are the Fresnel integrals.

For $\alpha \ll 1$ one has
\begin{eqnarray}
\Re \left[q_{jk} \right] &\approx& \delta_{kj} \left(1+\left(-\frac{1}{10}+\frac{-3+2 j^2 \pi ^2}{4 j^4
   \pi ^4}\right) \alpha ^2 + O(\alpha^4) \right) \nonumber \\
   &+&  (1-\delta_{kj}) \left[ \frac{8 (-1)^{j+k+1} j k \left(-12
   \left(j^2+k^2\right)+\left(j^2-k^2\right)^2 \pi^2\right) \alpha ^2}{\left(j^2-k^2\right)^4 \pi^4}   +O(\alpha^4)\right]\nonumber  \\
\Im \left[q_{jk} \right] &\approx& \delta_{kj} \left(\frac{1}{6} \left(2-\frac{3}{j^2 \pi ^2}\right)
   \alpha + O(\alpha^3) \right) +  (1-\delta_{kj}) \left[\frac{8 (-1)^{j+k} j k \alpha
   }{\left(j^2-k^2\right)^2 \pi ^2}  +O(\alpha^3)\right] \nonumber 
\end{eqnarray}

\subsection{Uniform compression}

In this case $\ell(t) = L_0 + a t$, where $a<0$ ($a>0$) corresponds to a compression (expansion) of the box.
Doescher and Rice have proved that for this case it is possible to obtain the solution to eq.~(\ref{eq_Sch}) exactly.
They assumed a particle in the ground state at the initial time:
\begin{eqnarray}
c_k(t_0) = \delta_{k1} \hspace{1cm} ; \hspace{1cm} d_k(t_0) = 0 \hspace{1cm} , \ \ k=1,2, \dots
\end{eqnarray}

Using eqs.~(\ref{eq_c}) and (\ref{eq_d}), and working with limited number of modes ($k_{\rm MAX} \approx 10$) 
we were able to reproduce the figures of \cite{Doescher69} quite accurately.

\begin{figure}[t!]
\begin{center}
\includegraphics[width=6cm]{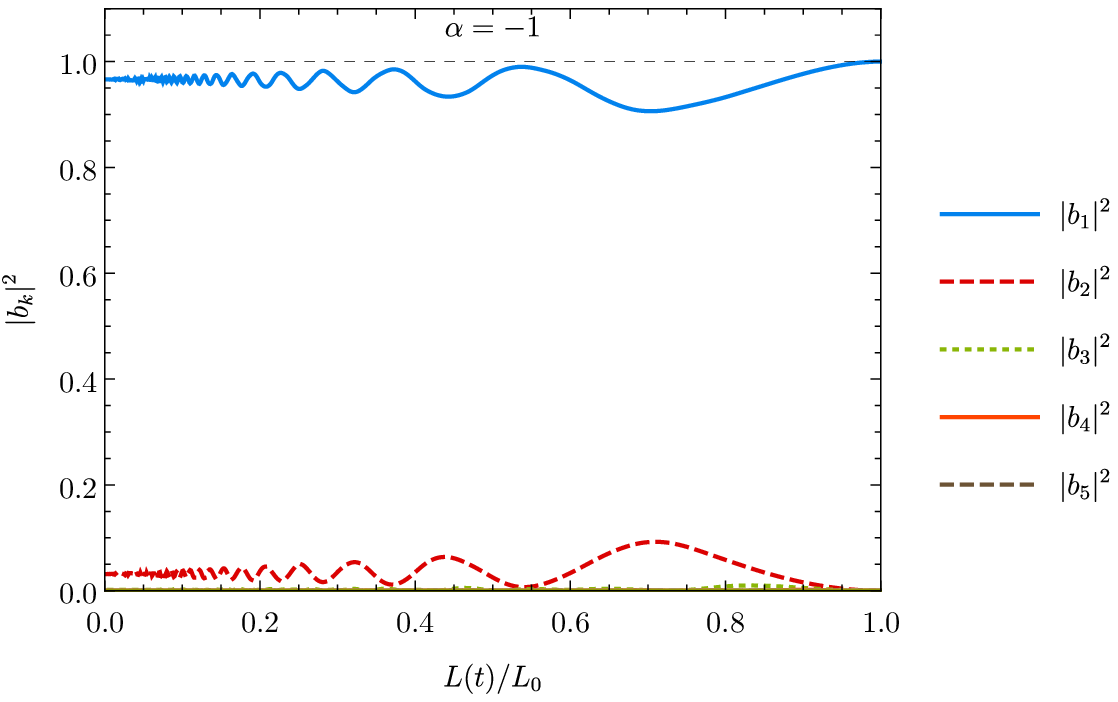}  \ \ 
\includegraphics[width=6cm]{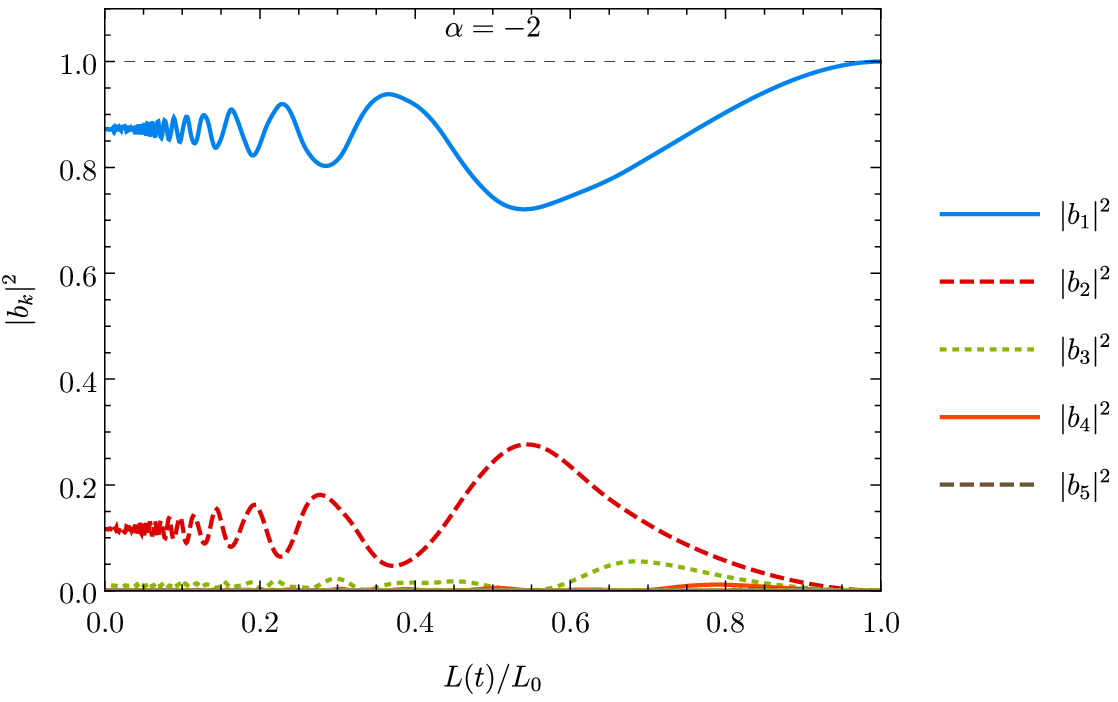} 
\caption{Analogous of Fig.2 (left plot) and Fig.3 (right plot) of Ref.~\cite{Doescher69}; }
\label{fig_1}
\end{center}
\end{figure}

\begin{figure}[h!]
\begin{center}
\includegraphics[width=6cm]{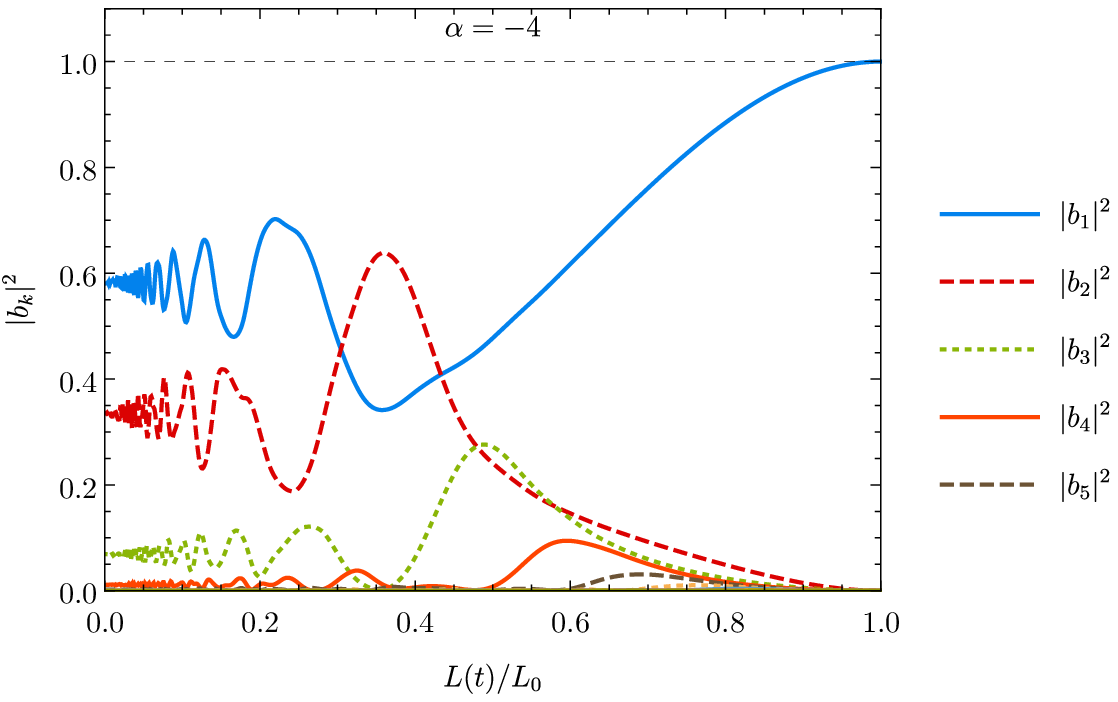}  \ \ 
\includegraphics[width=6cm]{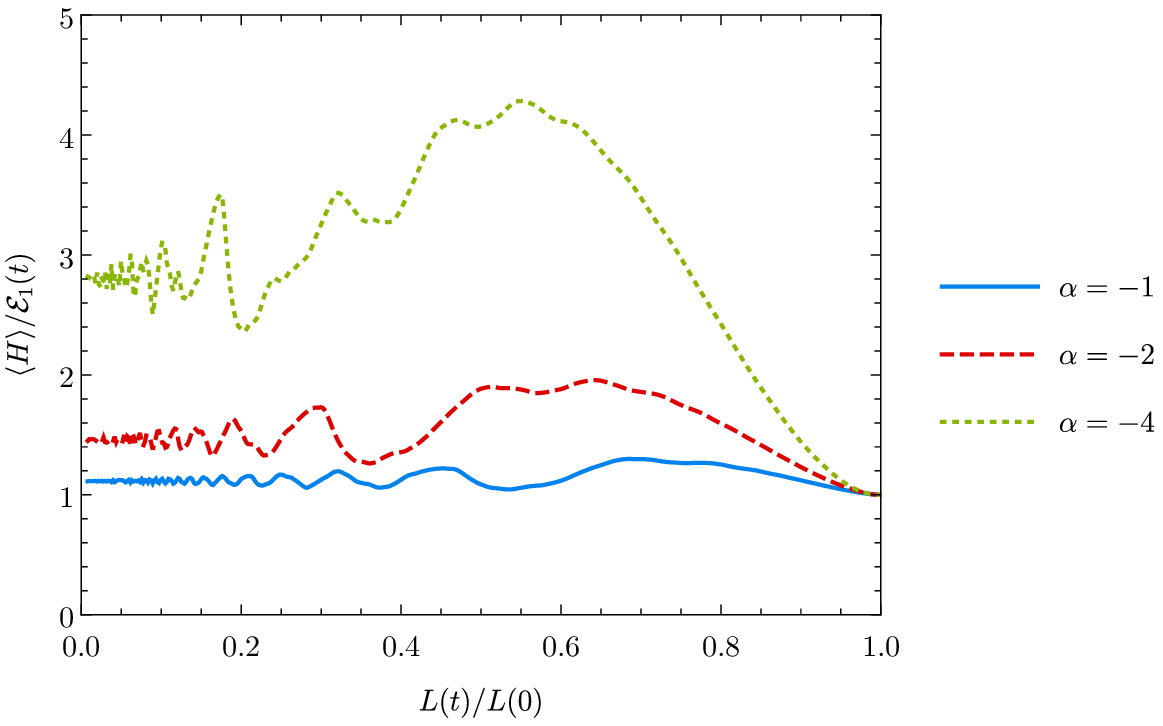} 
\caption{Analogous of Fig.4 (left plot) and Fig.5 (right plot) of Ref.~\cite{Doescher69}; }
\label{fig_2}
\end{center}
\end{figure}

\begin{figure}[h!]
\begin{center}
\includegraphics[width=12cm]{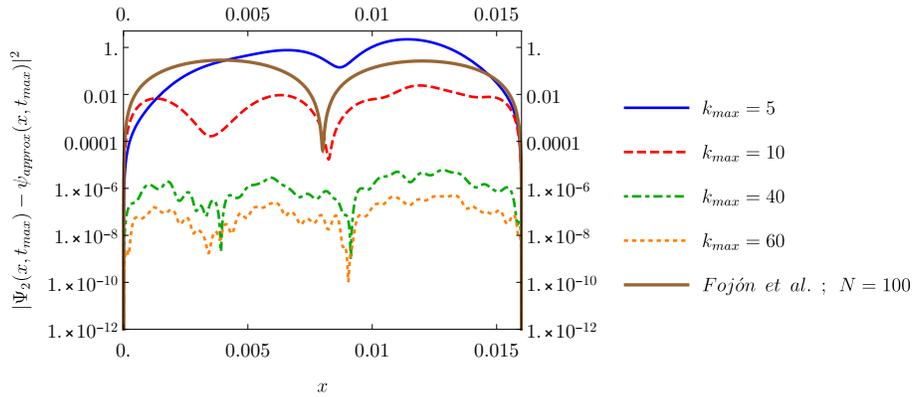}   
\caption{$\left| \Psi_2(x,t_{max}) -\psi_{approx}(x,t_{max}) \right|^2$ for the case of a wall 
moving with uniform velocity $a=-16$ at the time $t_{max} = 1/16-1/1000$, using the numerical solutions
to eq.~(\ref{eq_c}) and (\ref{eq_d}) with different number of modes (first four curves) and the numerical 
method of Foj\'on et al. with $N=100$ (last curve).
}
\label{fig_3}
\end{center}
\end{figure}

In Fig.~\ref{fig_3} we plot $\left| \Psi_2(x,t_{max}) -\psi_{approx}(x,t_{max}) \right|^2$ for the case of a wall 
moving with uniform velocity $a=-16$ at the time $t_{max} = 1/16-1/1000$, using the numerical solutions
to eq.~(\ref{eq_c}) and (\ref{eq_d}) with different number of modes (first four curves) and the numerical 
method of Foj\'on et al. with $N=100$ (last curve). Here $\Psi_2(x,t)$ is the exact solution of 
eq.~(\ref{eq_exact}) and $\psi_{approx}(x,t)$ is the approximate wave function obtained either with the spectral method or with the method of Foj\'on et al.

\begin{figure}[h!]
\begin{center}
\includegraphics[width=12cm]{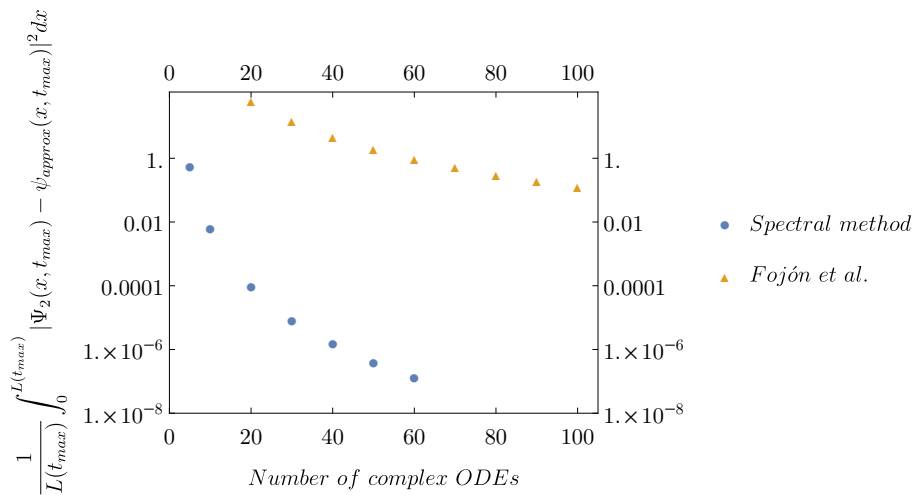}   
\caption{Average error, $\frac{1}{L(t_{max})} \int_0^{L(t_{max})} dx \left| \Psi_2(x,t_{max}) -\psi_{approx}(x,t_{max}) \right|^2$, for the case of Fig.~\ref{fig_3} using the spectral method or the method of Ref.~\cite{Fojon10}, 
as a function of the number of complex ODEs used.
}
\label{fig_4}
\end{center}
\end{figure}

Not surprisingly, the spectral method provides much smaller errors than the method of \cite{Fojon10},
despite having used a much finer discretization than in Ref.~\cite{Fojon10} ($N=100$).
Because the main source of error in this scheme comes from the spatial discretization, which
uses finite differences (FD), one should expect a slow decay of the error with the number of grid points used.
Fig.~\ref{fig_4} compares the average errors obtained using either the spectral method or the method of  \cite{Fojon10}, as functions of the number of complex ODEs.

\subsection{Oscillatory motion}

The main example discussed by Foj\'on et al. in \cite{Fojon10} is a box with a wall oscillating with the law
\begin{eqnarray}
\ell(t) = \ell(0) + a \sin \omega t 
\end{eqnarray}
where $\ell(0) = 1$. As mentioned before the initial wave function is the first excited state of the 
box with an uniformly moving wall at $t=0$.

The numerical results presented by these authors use $a = 0.3$ and $\omega = 1,10,4\pi^2$. For the case corresponding
to $\omega = 1$, one has $\alpha = 3/40 \ll 1$ and the initial wave function is well approximated by the first excited state
of the instantaneous energy  eigenfunctions. The situation is different for the cases $\omega = 10$ and $\omega = 4 \pi^2$,
which correspond to $\alpha = 3/4 \approx 1$ and $\alpha = 3\pi^2/10 \approx 3$, for which several eigenstates of the
instantaneous energy basis contribute.

\begin{figure}[t]
\begin{center}
\includegraphics[height=5cm]{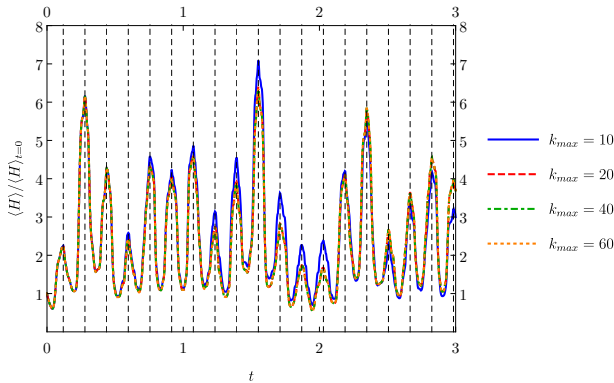} 
\caption{Mean value of the energy for the case of an oscillating wall with $a=0.3$ and $\omega = 4\pi^2$, obtained
using the spectral method with different number of modes. The energy is normalized with respect to the initial energy.
The thin dashed vertical lines correspond to multiple of the period of oscillation of the wall.}
\label{fig_7}
\end{center}
\end{figure}

In Fig.~\ref{fig_7} we display the mean value of the energy  for the case of an oscillating wall with $a=0.3$ and $\omega = 4\pi^2$, obtained
using the spectral method with different number of modes. The energy is normalized with respect to the initial energy.
As one can appreciate from the figure the results converge rather rapidly, with $20$ modes already giving an accurate description of the 
behavior.

\begin{figure}[t]
\begin{center}
\includegraphics[height=5cm]{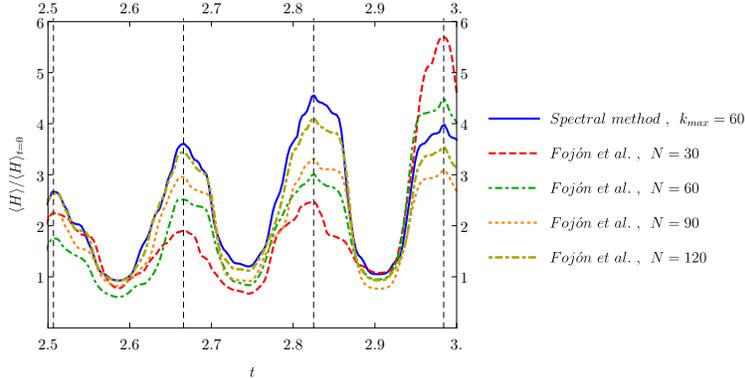} 
\caption{Mean value of the energy for the case of an oscillating wall with $a=0.3$ and $\omega = 4\pi^2$, obtained
using the method of Foj\'on et al. with different discretizations. The solid blue curve is the precise result obtained with
the spectral method with $60$ modes. The energy is normalized with respect to the initial energy.}
\label{fig_8}
\end{center}
\end{figure}

A similar plot, for $2.5 \leq t \leq 3$, is shown in Fig.~\ref{fig_8}, where now the method of Ref.~\cite{Fojon10} has been used, with different discretizations ($N=30,60,90,120$) and the results compared with the most precise results of Fig.~\ref{fig_7}, corresponding
to the spectral method with $k_{max}=60$. The case $N=30$ corresponds to the discretization used by Foj\'on et al. in \cite{Fojon10}
(check Figure 8 of that manuscript), and it provides a bad approximation to the true behavior of $\langle H \rangle$. 
Observe that even the results corresponding to the finest grid ($N=120$) are unsatisfactory for $t \approx 3$.

\begin{figure}[t]
\begin{center}
\includegraphics[height=5cm]{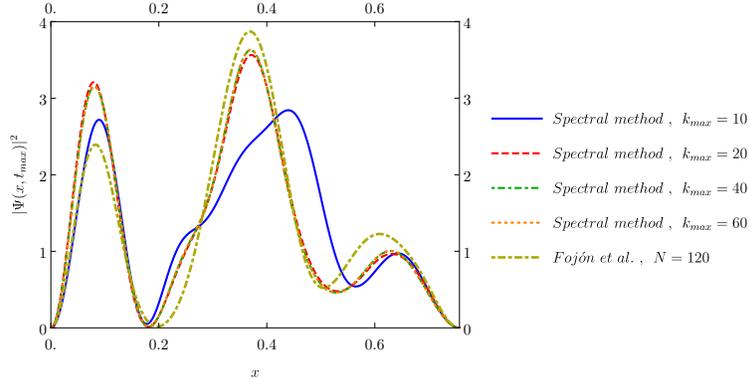} 
\caption{Probability density at the time $t_{max}=3$, for the case of an oscillating wall with $a=0.3$ and $\omega = 4\pi^2$, obtained
using the spectral method with $10,20,40,60$ modes and the method of Foj\'on et al. with $N=120$.}
\label{fig_9}
\end{center}
\end{figure}

Even if this was not done in Ref.~\cite{Fojon10}, it is useful to compare the probability density obtained with the different methods.
In Fig.~\ref{fig_9} we have plotted $|\Psi(x,t_{max})|^2$ obtained with the spectral method with $k_{max}=10,20,40,60$ modes and 
with the method of \cite{Fojon10}, with $N=120$. Consistent with what observed in Fig.~\ref{fig_7}, the results obtained with $k_{max}=20$ are
already quite precise; the results with the method of \cite{Fojon10}, with $N=120$, on the other hand, reproduce the general behavior 
only qualitatively.

\begin{figure}[t]
\begin{center}
\includegraphics[height=5cm]{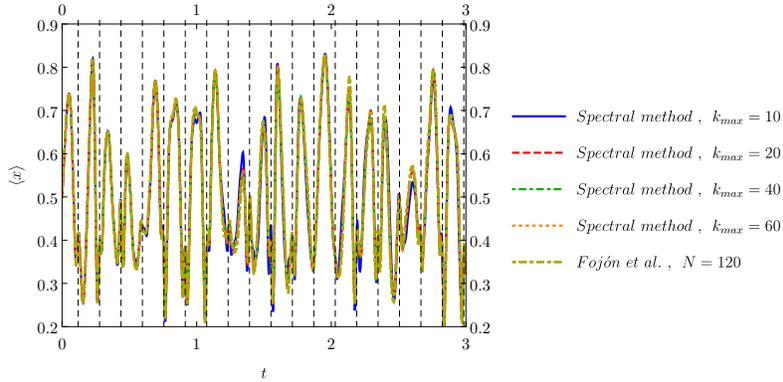} \\
\caption{Expectation value of the position obtained using the spectral method with different number of modes and the method of 
Ref.~\cite{Fojon10} with $N=120$.}
\label{fig_10}
\end{center}
\end{figure}

In Fig.~\ref{fig_10} we plot the mean value of the position obtained with the spectral method with $k_{max}=10,20,40,60$ modes and 
with the method of \cite{Fojon10}, with $N=120$ (compare with Fig.9 of Ref.~\cite{Fojon10}). In this case the method of  \cite{Fojon10} 
with $N=120$, reproduces rather well the spectral results.

\begin{figure}[t]
\begin{center}
\includegraphics[height=5cm]{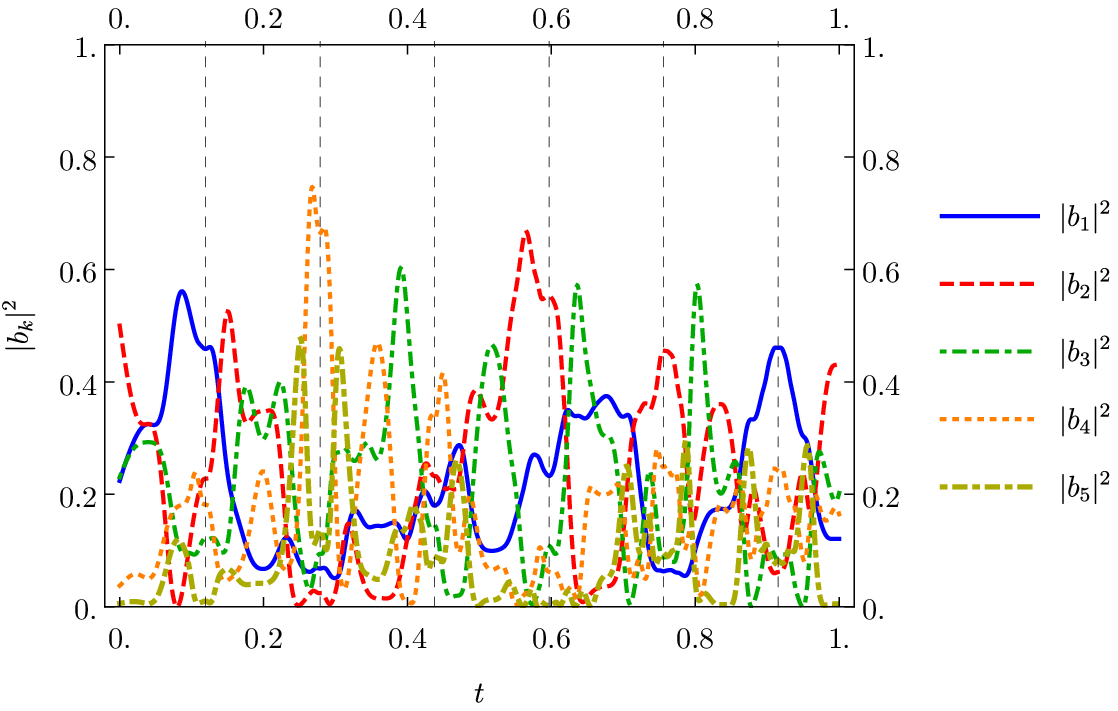} 
\caption{ $|b_k|^2(t)$ for $k=1,\dots 5$, as function of the time, for the case of an oscillating wall with 
$a=0.3$ and $\omega = 4\pi^2$, obtained using the spectral method with $60$ modes.}
\label{fig_11}
\end{center}
\end{figure}

In Fig.~\ref{fig_11} we show the expansion coefficients corresponding to the first $5$ modes of a 
system with an oscillating wall with $a=0.3$ and $\omega = 4\pi^2$.

\subsection{Sudden expansion}

The last example considered by Foj\'on et al. is the case of a "sudden expansion", in which the right wall moves according to the law
\begin{eqnarray}
\ell(t) = a - \frac{1}{1+b^2 t^2}
\end{eqnarray}

\begin{figure}[t]
\begin{center}
\includegraphics[height=5cm]{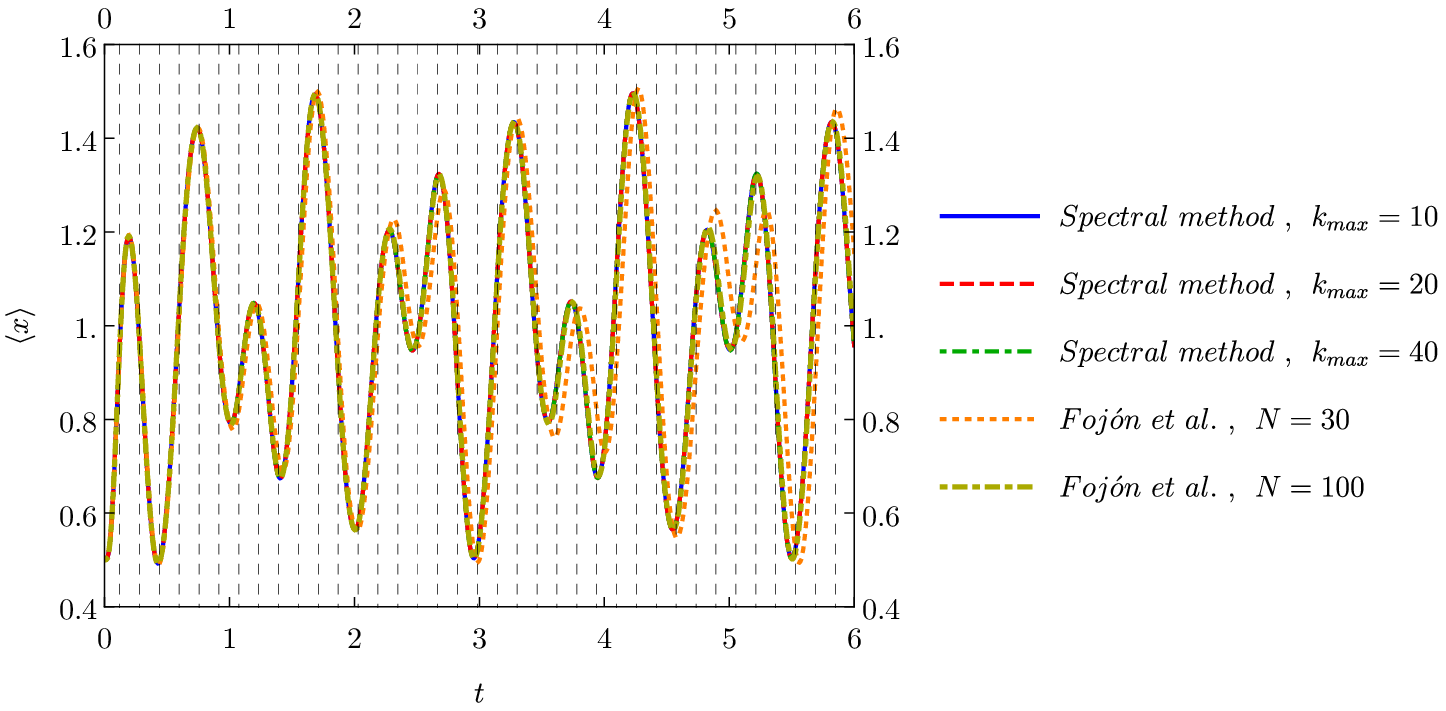} 
\caption{ Mean value of the position for the case of sudden expansion with $a=2$ and $b=10$, using the spectral method with $10$, $20$ and $40$ modes and the method of Foj\'on et al. with $N=30$ and $N=100$.}
\label{fig_12}
\end{center}
\end{figure}

These authors have considered the cases in which $a=2$ and $b=1,10,20$. Here we discuss the case of $b=10$, although similar considerations hold also for the remaining cases. In Fig.~\ref{fig_12} we plot the mean value of the position for the case of sudden expansion with $a=2$ and $b=10$, using the spectral method with $10$, $20$ and $40$ modes and the method of Foj\'on et at. with $N=30$ and $N=100$. In this case, we notice that the spectral method converges remarkably fast, with the curves corresponding 
to $10$, $20$ and $40$ superposing neatly. On the other hand, using the method of Foj\'on et al. with $N=30$ (i.e. the discretization 
used in Ref.~\cite{Fojon10}), one obtains a curve that is visibly different from the spectral curves; only using a much finer
grid, $N=100$, one is able to reproduce the results obtained with the spectral method with good accuracy.

\begin{figure}[t]
\begin{center}
\includegraphics[height=5cm]{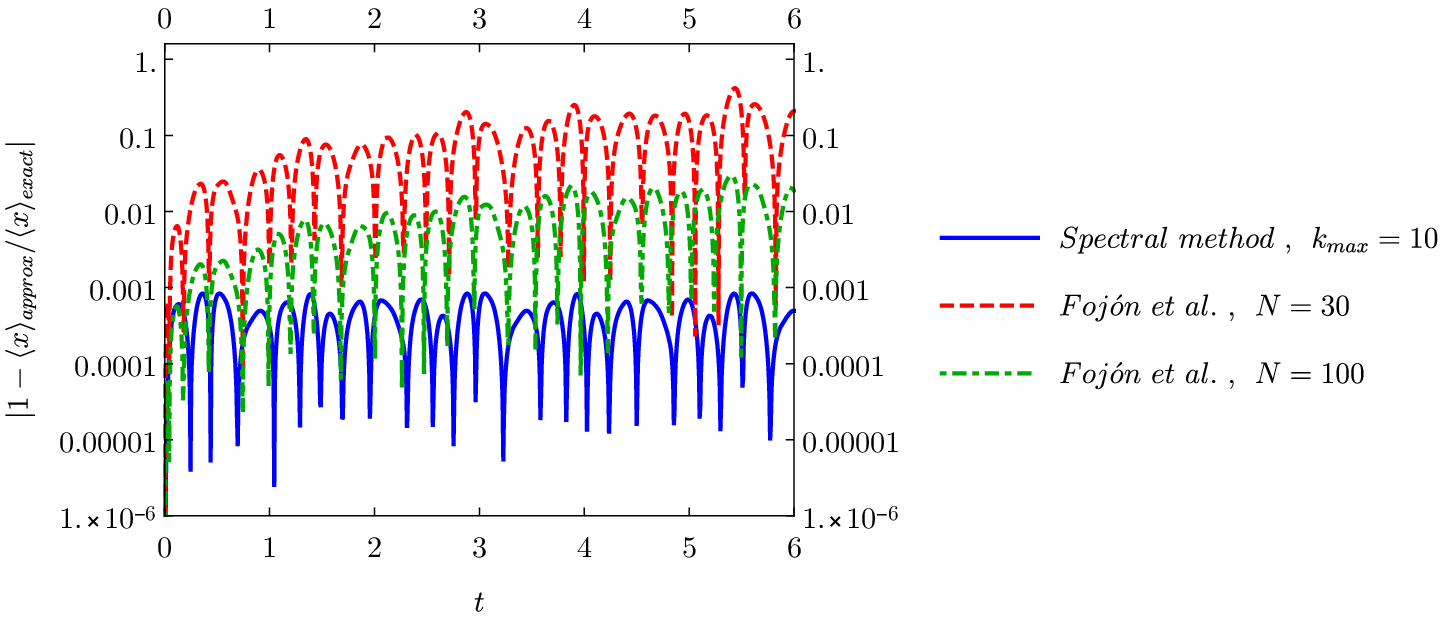} 
\caption{Error $|1-\langle x \rangle_{approx}/\langle x \rangle_{exact}|$ using the spectral method with $10$ modes and the method of Foj\'on et al. with $N=30$ and $N=100$. Here $\langle x \rangle_{exact}$ is approximated with the value obtained with the spectral method with $40$ modes.}
\label{fig_13}
\end{center}
\end{figure}

In Fig.~\ref{fig_13} we plot the error $\left| 1 - \langle x\rangle_{approx}/\langle x \rangle_{exact}\right|$, where 
$\langle x \rangle_{exact}$ is here approximated with the value obtained with the spectral method with $40$ modes. 
As one can easily appreciate the spectral method with just $10$ modes is more precise than the method of 
\cite{Fojon10} with $N=100$.

\section{Conclusions}
\label{sec:concl}

Many of the conclusions reached in Ref.~\cite{Fojon10} are misleading and in some cases incorrect. 
To start with, the authors claim the superiority of their method with respect to a spectral method, based on 
a decomposition of the initial wave function in the instantaneous energy basis. Although this claim could be 
discarded intuitively, since the equations used in the spectral method are {\sl exact} and the only approximation
enters in the definition of a cutoff mode and the subsequent numerical solution, we have performed precise 
numerical calculations, for all the cases treated in Ref.~\cite{Fojon10}, showing that the spectral method 
is far more precise than the method of \cite{Fojon10} and converging much more rapidly to the exact 
solution~\footnote{The slow convergence of the method of \cite{Fojon10} is due to the use of a simple finite difference approximation
in the approximation of the derivatives.}.

To support their claim of precision, these authors state three points, in their conclusions:
\begin{itemize}
\item That the numerical probability fluctuates within $3 \%$ in all cases considered by them;
\item That the numerical solution for the cases of a uniformly moving wall that they obtained with their method is
very accurate;
\item That they "compared the numerical solutions for different partitions of both space and time variables" and
that "the results remained essentially unchanged with different discretizations on the space and time variables".
\end{itemize}

Our observations on the points above are:
\begin{itemize}
\item The fact that the numerical probability fluctuates within $3-4 \%$ is not by itself a sufficient condition
for precision (actually, the probability obtained using the spectral method fluctuates within just 
$10^{-5}-10^{-6} \%$ for the same cases) : a better indicator of precision would be $\frac{1}{\ell(t)} \int_0^{\ell(t)} | \psi_{exact}(x,t) -  \psi_{approx}(x,t)|^2 dx$, which indeed is seen to be rather large even in the case of a uniformly moving wall using their method (see Fig.~\ref{fig_4});
\item A comparison of the numerical solutions for the case of a uniformly moving wall and of a sudden expansion using the method of \cite{Fojon10} and the spectral method, shows that the former is rather disappointing (the numerical results obtained with the method of 
\cite{Fojon10} with $N=100$ have much larger errors than the results obtained with the spectral method with just $10$ modes!);
\item The only way we can explain the observation of Foj\'on et al. is that they probably considered only discretizations
with slightly different numbers of points: in this case, due to the slow convergence of their method, the results would appear 
to change moderately. In our calculation with their method, we had to resort to much finer grids ($N \approx 100-120$) to obtain
results which would be qualitatively correct (at least for fast movements). Had the authors compared their method with
the spectral method, this problem would have been quite evident.
\end{itemize}

Another aspect of Ref.~\cite{Fojon10} that caught our attention is the distinction made therein between "standard"
and "nonstandard" regimes, meaning situations in which the wave function either preserves or not the number of nodes.
Although it is a surprise to the authors that the probability density has in some cases the form one would obtain 
as if the particle would find itself in an instantaneous energy eigenstate, this should not be a surprise at all: 
if the motion of the wall is sufficiently slow and the initial state is peaked around a mode, typically the coefficient
of the dominant mode oscillates with small amplitudes, while the remaining ones stay small. In this case
the solution can be well approximated  with the instantaneous wave function. On the other hand, when the wall moves fast, 
even if the particle starts in a given instantaneous energy eigenstate, several modes may become relevant at later times, thus
producing a very different wave function.

The discretization method will become increasingly more inefficient with the number of spatial dimensions. On the other hand, the spectral method is not expected to be dramatically affected.

\section*{Acknowledgements}
The research of P.A. was supported by Sistema Nacional de Investigadores (M\'exico).

\end{document}